\documentclass[10pt,conference]{IEEEtran}
\usepackage[utf8]{inputenc}

\usepackage{booktabs}
\usepackage{xcolor}
\usepackage{soul}
\usepackage{hyperref}
\usepackage{graphicx}
\usepackage{xspace}
\usepackage{cleveref}
\usepackage{pgfplots}
\usepackage{url}
\usepackage{balance}
\pgfplotsset{compat=1.8}
\usepgfplotslibrary{statistics}

\newcommand{\toolNameContent}{\textsc{RefactorInsight}}
\newcommand{\toolNameTitle}{\toolNameContent}
\newcommand{\toolName}{\emph{\toolNameContent}\xspace}
\usepackage{tikz}
\newcommand*\circled[1]{\tikz[baseline=(char.base)]{
            \node[shape=circle,draw,inner sep=2pt] (char) {#1};}}

\title{\toolNameTitle: Enhancing IDE Representation of Changes in Git with Refactorings Information}
\date{October 2020}

\author{
 \IEEEauthorblockN{Zarina Kurbatova,\IEEEauthorrefmark{1} Vladimir Kovalenko,\IEEEauthorrefmark{1}\IEEEauthorrefmark{2} Ioana Savu,\IEEEauthorrefmark{3} Bob Brockbernd,\IEEEauthorrefmark{3} Dan Andreescu,\IEEEauthorrefmark{3} \\ Matei Anton,\IEEEauthorrefmark{3} Roman Venediktov,\IEEEauthorrefmark{4} Elena Tikhomirova,\IEEEauthorrefmark{1} Timofey Bryksin\IEEEauthorrefmark{1}\IEEEauthorrefmark{4}}
    \IEEEauthorblockA{\IEEEauthorrefmark{1}\textit{JetBrains Research},  \IEEEauthorrefmark{2}\textit{JetBrains N.V.},
    \IEEEauthorrefmark{3}\textit{Delft University of Technology},
    \IEEEauthorrefmark{4}\textit{Higher School of Economics}}
    \IEEEauthorblockA{\{zarina.kurbatova, vladimir.kovalenko, elena.tikhomirova, timofey.bryksin\}@jetbrains.com,\\ \{a.i.savu-1, B.J.Brockbernd, D.Andreescu, M.Anton\}@student.tudelft.nl, \{rmvenediktov\}@edu.hse.ru}
}

\begin{document}

\maketitle

\begin{abstract}
Inspection of code changes is a time-consuming task that constitutes a big part of everyday work of software engineers.
Existing IDEs provide little information about the semantics of code changes within the file editor view. Therefore developers have to track changes across multiple files, which is a hard task with large codebases.

In this paper, we present \toolName, a plugin for IntelliJ IDEA that introduces a smart diff for code changes in Java and Kotlin where refactorings are auto-folded and provided with their description, thus allowing users to focus on changes that modify the code behavior like bug fixes and new features.
\toolName supports three usage scenarios: viewing smart diffs with auto-folded refactorings and hints, inspecting refactorings in pull requests and in any specific commit in the project change history, and exploring the refactoring history of methods and classes.
The evaluation shows that commit processing time is acceptable: on median it is less than 0.2 seconds, which delay does not disrupt developers' IDE workflows.

\toolName is available at \url{https://github.com/JetBrains-Research/RefactorInsight}. The demonstration video is available at \url{https://youtu.be/-6L2AKQ66nA}.
\end{abstract}

\section{Introduction}
Inspecting code changes is a daily routine for software engineers.
Improving this procedure and its efficiency has been a focus of many efforts ~\cite{tao2012software, bacchelli2013expectations, sadowski2018modern}.
For example, developers examine changes made by their colleagues to search for possible defects, check if the code follows project conventions, suggest better solutions and share the knowledge with the team~\cite{sadowski2018modern}.
Manual inspection of a large number of changes may be challenging. 
To reduce the cognitive load on developers, it could be useful to simplify the process of inspecting changes by presenting them in a more structured and compact way. 

A lot of features in modern IDEs are designed to boost developers' productivity. 
For example, code completion helps developers to write code faster and make fewer errors through reduced typing, while code folding\footnote{Code folding in IntelliJ IDEA: \url{https://www.jetbrains.com/help/idea/working-with-source-code.html\#code_folding}} helps developers to maintain focus by collapsing irrelevant code elements. 
Some of such productivity-enhancing features are related to the process of reviewing code changes.
For example, IntelliJ IDEA allows its users to explore the change history of files or pieces of code\footnote{Explore change history in IntelliJ IDEA: \url{https://www.jetbrains.com/help/idea/viewing-changes-information.html\#changes_history}}.

However, currently IDEs do not provide much insight into the semantics of changes, and developers still need to put effort into figuring out whether and how the changes influence the behavior of code and whether changes across multiple files are connected to each other.
Visualizing information related to the semantics of code changes is a promising way to augment the presentation of changes in an IDE.
One way of doing it is to separate the changes that represent refactorings from such changes that modify the behavior of code.
Refactoring is a fundamental practice in software development and consists in changing code without affecting its behavior~\cite{fowler2018refactoring}. 
Leppanen et al. interviewed software developers about the role and importance of refactoring~\cite{leppanen2015refactoring}. 
The authors show that developers are cautious of introducing new defects during refactoring and emphasize the importance of careful checking of such kind of changes. 
These findings suggest that separating refactorings from other changes could be a promising developer assistance technique.

Several refactoring-aware tools have been proposed to help developers during the inspection of code changes.
Some of the tools are implemented as plugins for the Eclipse IDE and, sadly, are limited in the number of refactorings they are able to detect~\cite{alves2017refactoring, ge2017refactoring}.
Other tools are implemented as plugins to the Google Chrome browser and provide information about performed refactorings in changes on GitHub.
RAID~\cite{brito2021raid} can detect only 13 refactoring types in commits and pull requests, while Refactoring Aware Commit Review works~\cite{refactoringaware-chrome} only with commits.

We present \toolName, a plugin for IntelliJ IDEA~\cite{intellij-idea} that auto folds refactorings in code diffs in Java and Kotlin and shows hints with their short descriptions.
Also the plugin allows users to view a list of performed refactorings in any specific commit or pull request as well as the refactoring history for methods and classes.
To detect refactorings, our tool relies on RefactoringMiner~\cite{Tsantalis:TSE:2020:RefactoringMiner2.0}, which demonstrates the highest accuracy among existing approaches as shown in~\cite{Tsantalis:TSE:2020:RefactoringMiner2.0}.
\toolName is published in the JetBrains plugin repository\footnote{\toolName in the plugin repository:  \url{https://plugins.jetbrains.com/plugin/14704-refactorinsight}}. 
Its source code is available on GitHub\footnote{\toolName on GitHub: { https://github.com/JetBrains-Research/RefactorInsight}}. 

\section{Background}

Existing research suggests that developers benefit from refactoring-aware tool support in code review. 
Kim et al.~\cite{kim2012field} interviewed developers from Microsoft about the challenges and benefits of refactorings. 
One of the findings of the study is that refactoring may entail merge conflicts, regression bugs, impeded code review, and other risks.
Besides, the developers reported that changes that contain refactorings are more difficult to review.
Thus, the authors conclude that development of tools that make inspection of refactorings isolated from other changes can be a promising direction for future research.
Alomar et al.~\cite{alomar2021refactoring} conducted an industrial case study involving developers from Xerox to investigate the mechanisms of refactoring-aware code review.
The results show that one of the main challenges is that refactoring changes are interleaved with bug fixes and new features.
The reviewers need to make an effort to separate the changes in code logic and behavior-preserving changes.

There have been proposed several tools for automatic detection of refactorings in code changes~\cite{Tsantalis:TSE:2020:RefactoringMiner2.0, silva2017refdiff}. 
Tsantalis et al. presented a standalone library called RefactoringMiner~\cite{Tsantalis:TSE:2020:RefactoringMiner2.0}. It demonstrates 99.6\% precision and 94\% recall, which are the highest results among other approaches, as shown in its evaluation. The tool supports the detection of 40 refactoring types. It analyzes the commit history using a statement matching algorithm and refactoring detection rules, and provides a list of refactorings with descriptions for each commit. 
Each description contains information about code elements involved in the refactoring and their code ranges before and after the refactoring.
RefactoringMiner's output is programmatic, but, if integrated into existing tools that support working with Git, can augment diffs and history, significantly saving developers' code inspection time. RefactoringMiner is also limited to working with Java projects only. We reduced both usability limitations, as we show in Section~\ref{sec:implementation}. 

There is a number of working plugins for Eclipse IDE~\cite{ge2017refactoring, matsuda2015hierarchical, zhang2015interactive, alves2017refactoring} and extensions for Chrome which visualize refactoring information in Git diffs. Let's have a brief overview of their functionality.

Ge et al. presented a refactoring-aware plugin for Eclipse called ReviewFactor~\cite{ge2017refactoring} that separates refactoring and non-refactoring changes.
The tool is able to detect merely five refactoring types: Rename Type, Rename Method, Move Method, Extract Method, and Inline Method.
A case study of ReviewFactor shows that the separation of changes can simplify the code review process.
Alves et al. presented RefDistiller~\cite{alves2017refactoring}, a plugin for Eclipse that detects potential refactoring anomalies that might occur when developers' edits lead to subsequent behavioral changes after code refactoring, which is a behavior-preserving edit by definition.
The tool is able to capture six refactoring types, namely, Extract Method, Inline Method, Move Method, Pull Up Method, Push Down Method, and Rename Method.

Brito et al. presented an extension for Google Chrome called RAID~\cite{brito2021raid} that supplements code diffs in pull requests on GitHub with information about performed refactorings.
To detect refactorings in code changes, RAID relies on RefDiff~\cite{silva2017refdiff} that currently supports the detection of 13 refactoring types in code changes in Java, C, JavaScript, and Go.
The tool does not detect low-level refactorings that relate to variables.
Another extension for Google Chrome called Refactoring Aware Commit Review~\cite{refactoringaware-chrome} relies on the above-mentioned RefactoringMiner library~\cite{Tsantalis:TSE:2020:RefactoringMiner2.0} to detect refactorings and highlights them in commits on GitHub.
The extension works only with commits and as well as RAID works only in the Google Chrome browser.

\begin{table}
    \centering
        \caption{Refactoring types supported by \toolName.}
    \label{tab:refactoringTypes}
    \begin{tabular}{ll}
    \toprule
    \multicolumn{1}{c}{\textbf{Java, Kotlin}} & \multicolumn{1}{c}{\textbf{Java}} \\ \midrule
    Rename Variable/Method/Class & Merge Variable/Parameter/Field\\
    Rename Parameter/Field & Change Package\\
    Extract Variable/Method/Class & Parameterize Variable \\ 
    Extract Interface/Superclass & Extract Subclass \\ 
    Extract And Move Method & Replace Variable/Field with Field \\
    Move Method/Class &  Split Variable/Parameter/Field \\
    Move And Rename Method & Move And Rename Field\\ 
    Pull Up Method/Field & Pull Up Field\\
    Push Down Method/Field &  Push Down Field\\
    Inline Method &  \\
    Move And Inline Method & \\
    Change Variable/Parameter & \\
    Change Field Type & \\
    Add/Remove/Reorder Parameter & \\
    \bottomrule
\end{tabular}
\end{table}

\section{Implementation}
\label{sec:implementation}
Some users prefer to review pull requests from GitHub within their IDE because it minimizes the context switches between the IDE and the browser and therefore saves time~\cite{pull-requests-ticket}.
Our solution is implemented as an IntelliJ IDEA plugin that processes pull requests and the project Git history and integrates refactoring reports in file  diffs.
\toolName detects 40 types of refactorings in Java code changes and 19 types of refactorings in Kotlin code changes.
To save processing time, all information about refactorings detected in a project is stored locally.

\subsection{Workflows}

\subsubsection{Viewing diffs with folded refactorings}

\begin{figure}
  \centering
  \includegraphics[width=\columnwidth]{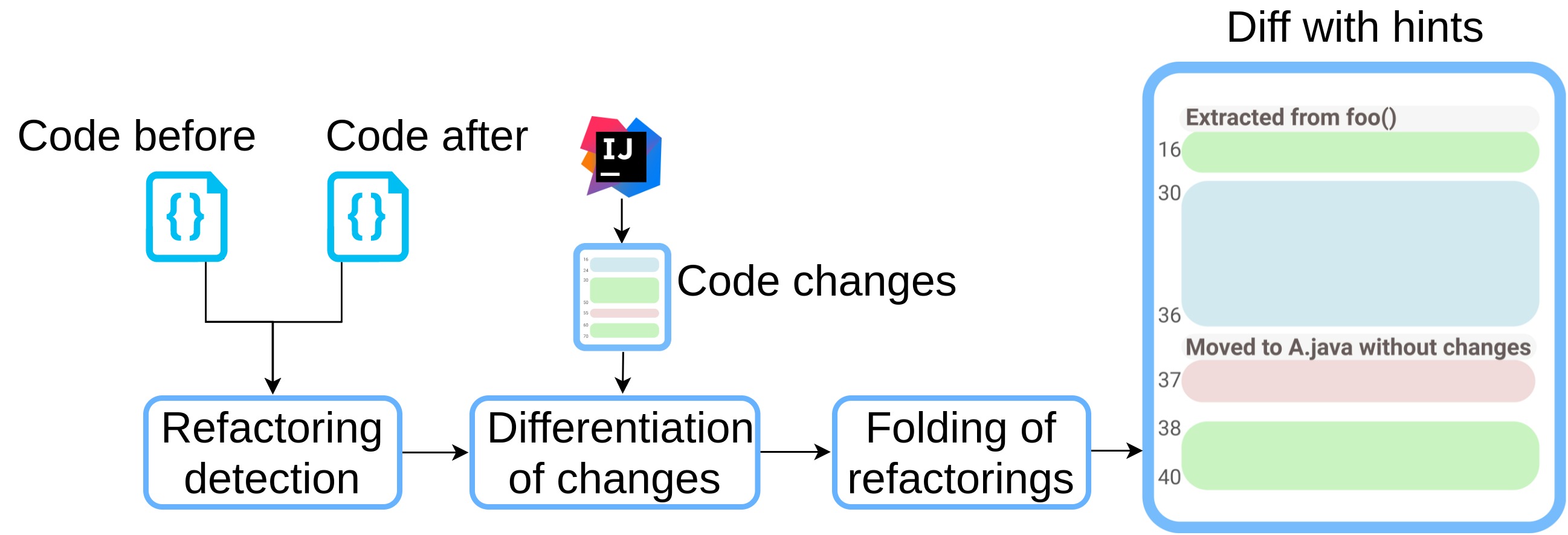}
  \caption{Overview of \toolName.}
  \label{fig:toolWorkflow}
\end{figure}

\begin{figure}
\centering
  \includegraphics[width=0.5\textwidth]{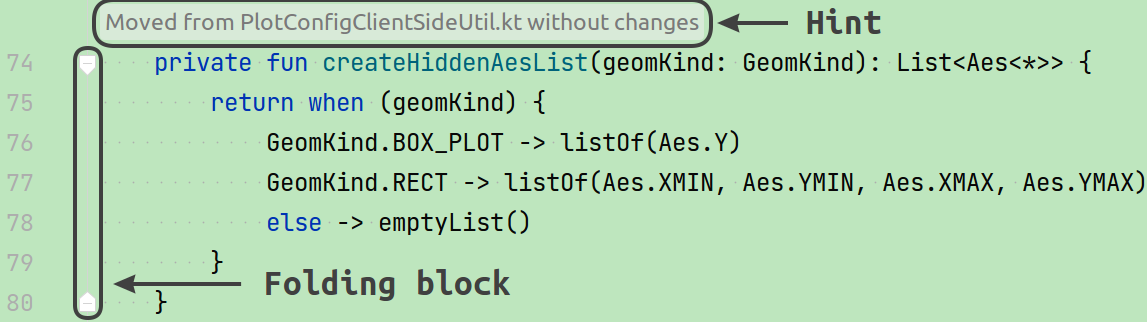}
  \caption{Example of a hint shown in an expanded diff. The hint explains that the newly added method was moved from another existing class and clarifies that there were no changes.}
  \label{fig:hintExample}
\end{figure}

The primary workflow of \toolName is presented in~\Cref{fig:toolWorkflow}.
The plugin detects refactorings performed in the corresponding commit.
Next, using the IntelliJ Platform, the plugin gets code changes in the selected file.
Then, the plugin differentiates the changes as refactoring and non-refactoring ones to decide which of them to fold.
Finally, the plugin shows a diff window with auto-folded refactorings and hints with short descriptions of folded refactorings.
An example of a hint is provided in~\Cref{fig:hintExample}.

Auto-folding and annotating refactorings allows users to focus on new features or bug fixes in the diff. 
When necessary, refactored code can be viewed in a number of ways.
The user can:
\begin{itemize}
 \item Expand the folded lines to view them;
 \item Read the hint describing where the method was moved to or from (for Move Method, Pull Up Method, and Push Down Method refactorings), or that a piece of code was extracted into a new method from another existing method (Extract Method refactoring), or that a method was inlined into another one (Inline Method refactoring).
\end{itemize}

\subsubsection{Viewing the list of refactorings per commit or pull request}
\begin{figure}
  \centering
  \includegraphics[width=\columnwidth]{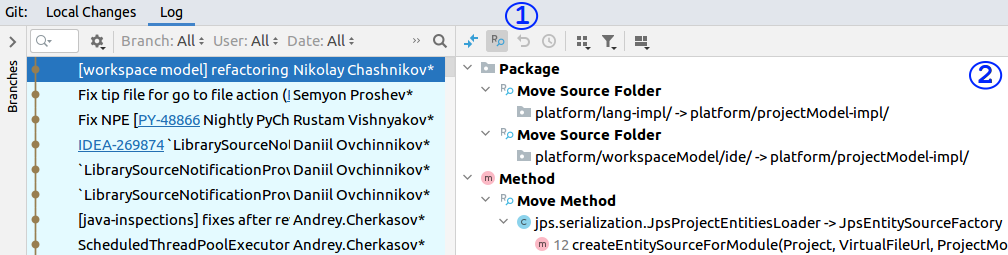}
  \caption{Example of the list of detected refactorings that opens when the user clicks a commit on the Git Log tab. The plugin extends the tab with the button \protect\circled{1} and the panel with the listed refactorings \protect\circled{2}.}
  \label{fig:gitLogTab}
\end{figure}

\toolName allows users to check the list of performed refactorings in any specific commit on the Git Log tab or in any pull request. 
The plugin shows a list of refactorings with short descriptions. For convenience, the refactorings in the list are grouped by the element's level: package, class, method, and variable.
An example list of refactorings per commit is provided in~\Cref{fig:gitLogTab}.

When the user double-clicks a list item, \toolName displays the refactored code in a separate diff window.
\toolName notes the ``before'' and ``after'' location of the refactored piece and shows code related only to the refactoring.

\subsubsection{Viewing the history of refactorings for methods and classes}
When the user right-clicks a code element in the editor and selects ``Show Refactoring History'', the IDE displays a list of refactorings this element was involved in.
This option is only available if any refactorings have been detected in the history of changes for this element.
To mine refactorings in the whole commit history of the project, the user should select ``Mine All Refactorings'' on the Tools tab in the IDE.

\subsection{Refactoring detection}
\toolName detects refactorings in Java and Kotlin changes.
To identify refactorings in Java, we use RefactoringMiner 2.0 which works only with Java projects.
We overcome this limitation by extending RefactoringMiner to Kotlin in our library called kotlinRMiner that currently supports the detection of 19 refactoring types\footnote{kotlinRMiner: \url{https://github.com/JetBrains-Research/kotlinRMiner}}. 
kotlinRMiner is based on the RefactoringMiner's heuristics and refactoring detection rules. It relies on the Kotlin compiler to parse the code and extract its inner representation.
The full list of refactoring types supported in \toolName is presented in~\Cref{tab:refactoringTypes}.
Both tools receive a commit hash and project location as input and return a list of detected refactorings with descriptions containing the refactoring type, names of involved elements, and their code ranges.

\subsection{Integration with the IDE}
\toolName differentiates the changes as refactoring and non-refactoring ones.
When the user selects a file in a commit, \toolName augments the diff window with information about refactorings, using the IntelliJ Platform\footnote{IntelliJ Platform: \url{https://www.jetbrains.com/opensource/idea/}} API which allows users to develop custom code inspections, implement tool windows, and add custom actions and listeners.
To differentiate the changes, the plugin uses the information provided by RefactoringMiner and kotlinRMiner and searches for the code elements involved in the refactoring across the full list of changes.
To check if a method was moved with or without changes, the plugin compares the method's body before and after the commit.
To retrieve information about commits and pull requests, we use the Git4Idea\footnote{Git4Idea: \url{https://github.com/JetBrains/intellij-community/tree/master/plugins/git4idea}} module of the IntelliJ Platform, which is responsible for Git integration.

\subsection{Folding of refactorings}
IntelliJ IDEA shows a standard diff window, with one pane showing the old version of the file with removed and modified lines highlighted, and the other displaying the new version of the file, with added and modified lines highlighted.
We extend the behavior of the diff window by overriding the method that is called each time when the user clicks a file to check its diff.

The plugin adds folding blocks to both panes in the diff window by changing its inlay model, an entity used in the IntelliJ Platform to introduce some custom visual elements into the IDEs.
The folding is used to collapse the lines and thus reduce the amount of code the developer needs to scan initially.
To find out the borders of each folding block, the plugin calculates the text offsets of elements involved in the refactoring. 
\toolName folds the refactorings for methods because they usually consist of at least several code lines.
In case of the Move Method refactoring the plugin also extends the hint with information on whether the method was changed during the move.

\subsection{Visualization}
The way the plugin visualizes each refactoring depends on its type and the number of involved elements. 

First, \toolName processes the information provided by RefactoringMiner and kotlinRMiner---the type of refactoring and the ranges of all involved elements---and calculates the number of panes required to visualize it.
For example, in case of the Move Method refactoring the plugin displays a diff window with two panes: one with the location where the method was moved from, and the other where the method was moved to.

After that, \toolName combines related refactorings.
For example, for the Extract Superclass refactoring, it groups the refactorings that introduce the same new parent class. 
It improves the readability by showing related changes from several files in one place, which is especially useful in case of a large number of changes.

Finally, when the user right-clicks a commit in the list of commits, \toolName shows the diff window, highlighting modified code with blue color and added code with green color (for example, green is used to highlight a new parameter introduced in the Add Parameter refactoring).

\section{Evaluation}
Commit processing by the plugin should not cause extra delay in developers' work with the project history in the IDE.
To evaluate the execution time of \toolName, we selected two Kotlin and two Java open-source projects from GitHub that had at least ten contributors, a permission license, and commits after 1 May 2020.
The selected projects are presented in~\Cref{tab:evaluation-projects}. 

Then we employed \toolName to discover refactorings in each commit in project histories and measured the time it needed to run RefactoringMiner and kotlinRMiner sequentially and process the results using the \texttt{System.nanoTime} Java method.
We ran \toolName five times on each commit in each project and then calculated the average processing time for each commit to get more comprehensive results.
In total, \toolName processed 960 commits from four projects and detected 1,811 refactorings in them.

\begin{table}[h!]
    \centering
        \caption{Commits per project processed by \toolName.}
    \label{tab:evaluation-projects} 
    \begin{tabular}{r r r r} 
 \toprule
 \textbf{Name} & \textbf{Language} & \textbf{\#Commits} & \textbf{\#Refactorings found}\\ [0.5ex] 
 \midrule
cola & Java & 196 & 1,212\\
jcasbin & Java & 305 & 315\\
mavericks & Kotlin & 304 & 183\\
khttp & Kotlin & 155 & 101\\ 
\midrule
\textbf{Total} & & 960 & 1,811\\ 
 \bottomrule
 \end{tabular}
\end{table}

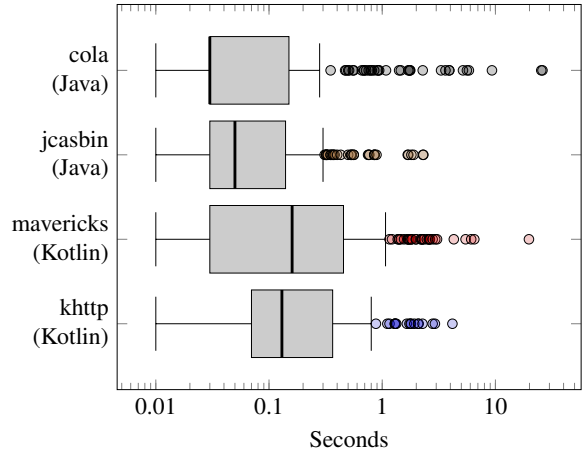
\begin{figure}
\begin{tikzpicture}[scale=0.9]
  \begin{axis}
    [
    ytick={1,2,3,4},
    yticklabels={khttp \\(Kotlin),
    mavericks \\(Kotlin),
    jcasbin \\(Java),
    cola \\(Java)},
    yticklabel style={align=right},
    xmode=log,
    xticklabels={0, 0.01, 0.1, 1, 10},
    xlabel={Seconds},
    every axis plot/.append style={fill,fill opacity=0.2},
    boxplot/every median/.style={black,very thick,solid}
    ]
       \addplot + [
  mark = *,
  boxplot,
  black
  ]
  table [row sep = \\, y index = 0] {
    data \\
0.03 \\ 0.61 \\ 1.11 \\ 0.1 \\ 0.08 \\ 0.02 \\ 0.1 \\ 0.17 \\ 0.11 \\ 2.07 \\ 0.24 \\ 0.6 \\ 0.01 \\ 0.1 \\ 0.88 \\ 2.79 \\ 0.09 \\ 0.17 \\ 0.03 \\ 0.41 \\ 0.16 \\ 0.03 \\ 0.14 \\ 0.13 \\ 0.18 \\ 0.12 \\ 0.02 \\ 0.11 \\ 0.01 \\ 0.23 \\ 0.34 \\ 0.14 \\ 0.02 \\ 0.22 \\ 0.11 \\ 0.37 \\ 0.41 \\ 0.48 \\ 2.27 \\ 0.01 \\ 0.43 \\ 1.92 \\ 0.41 \\ 0.21 \\ 0.05 \\ 0.53 \\ 0.1 \\ 0.14 \\ 0.11 \\ 0.12 \\ 0.11 \\ 0.11 \\ 0.02 \\ 0.1 \\ 2.92 \\ 0.02 \\ 0.01 \\ 0.02 \\ 0.13 \\ 1.74 \\ 0.1 \\ 0.11 \\ 0.01 \\ 0.02 \\ 0.35 \\ 0.14 \\ 0.21 \\ 0.02 \\ 0.47 \\ 1.3 \\ 0.02 \\ 0.12 \\ 0.38 \\ 0.79 \\ 0.19 \\ 0.13 \\ 0.2 \\ 0.22 \\ 0.09 \\ 1.16 \\ 0.02 \\ 0.12 \\ 0.62 \\ 0.09 \\ 0.02 \\ 0.15 \\ 0.09 \\ 0.08 \\ 0.03 \\ 0.02 \\ 0.03 \\ 0.03 \\ 0.22 \\ 0.13 \\ 0.28 \\ 1.79 \\ 0.75 \\ 0.28 \\ 0.02 \\ 0.01 \\ 0.08 \\ 0.19 \\ 0.08 \\ 1.65 \\ 0.02 \\ 0.36 \\ 1.33 \\ 0.09 \\ 0.12 \\ 0.02 \\ 0.02 \\ 0.56 \\ 0.17 \\ 0.18 \\ 0.02 \\ 0.12 \\ 0.02 \\ 0.13 \\ 0.32 \\ 0.19 \\ 0.02 \\ 0.19 \\ 0.8 \\ 2.1 \\ 0.04 \\ 0.03 \\ 0.5 \\ 0.03 \\ 0.8 \\ 0.04 \\ 1.3 \\ 0.08 \\ 0.13 \\ 0.17 \\ 0.12 \\ 0.02 \\ 0.17 \\ 0.26 \\ 0.17 \\ 0.46 \\ 0.12 \\ 1.29 \\ 0.64 \\ 0.09 \\ 0.13 \\ 0.2 \\ 4.17 \\ 0.11 \\ 1.79 \\ 0.19 \\ 0.06 \\ 0.54 \\ 0.17 \\ 0.3 \\ 0.19 \\
};

       \addplot + [
  mark = *,
  boxplot,
  black
  ]
  table [row sep = \\, y index = 0] {
    data \\
0.24 \\ 0.18 \\ 0.03 \\ 0.12 \\ 0.69 \\ 0.76 \\ 1.06 \\ 0.02 \\ 0.02 \\ 0.75 \\ 0.03 \\ 0.03 \\ 0.09 \\ 0.58 \\ 0.02 \\ 0.04 \\ 0.03 \\ 0.03 \\ 0.03 \\ 1.46 \\ 0.04 \\ 19.78 \\ 0.05 \\ 0.21 \\ 0.01 \\ 0.05 \\ 0.02 \\ 0.47 \\ 0.65 \\ 1.03 \\ 0.02 \\ 0.75 \\ 0.48 \\ 0.19 \\ 0.02 \\ 0.02 \\ 0.02 \\ 0.29 \\ 0.46 \\ 4.29 \\ 0.63 \\ 0.18 \\ 0.11 \\ 0.03 \\ 0.04 \\ 0.01 \\ 0.02 \\ 0.41 \\ 2.58 \\ 0.02 \\ 0.26 \\ 0.49 \\ 0.22 \\ 0.09 \\ 0.18 \\ 0.04 \\ 2.48 \\ 1.98 \\ 0.03 \\ 0.31 \\ 0.12 \\ 1.67 \\ 0.03 \\ 2.63 \\ 0.24 \\ 0.02 \\ 3.05 \\ 0.04 \\ 0.13 \\ 0.02 \\ 0.02 \\ 0.04 \\ 0.58 \\ 0.03 \\ 0.35 \\ 0.12 \\ 0.16 \\ 0.02 \\ 0.02 \\ 0.03 \\ 0.03 \\ 2.34 \\ 0.25 \\ 1.56 \\ 6.11 \\ 0.12 \\ 0.7 \\ 0.1 \\ 0.02 \\ 0.15 \\ 0.24 \\ 0.02 \\ 2.27 \\ 0.26 \\ 0.8 \\ 0.1 \\ 0.01 \\ 0.17 \\ 0.02 \\ 2.79 \\ 0.03 \\ 0.23 \\ 0.01 \\ 0.26 \\ 1.7 \\ 0.02 \\ 0.04 \\ 1.73 \\ 0.17 \\ 0.28 \\ 0.09 \\ 0.03 \\ 0.02 \\ 0.3 \\ 0.16 \\ 0.02 \\ 0.9 \\ 0.09 \\ 0.36 \\ 0.16 \\ 0.18 \\ 0.43 \\ 0.54 \\ 0.42 \\ 2.23 \\ 0.09 \\ 0.17 \\ 1.03 \\ 0.19 \\ 0.74 \\ 0.1 \\ 2.94 \\ 0.04 \\ 0.3 \\ 0.55 \\ 0.02 \\ 0.55 \\ 0.14 \\ 0.23 \\ 1.05 \\ 0.1 \\ 0.17 \\ 0.02 \\ 0.4 \\ 1.22 \\ 1.7 \\ 0.17 \\ 1.77 \\ 0.03 \\ 0.17 \\ 0.22 \\ 0.04 \\ 0.02 \\ 0.66 \\ 0.03 \\ 0.02 \\ 0.29 \\ 0.02 \\ 0.28 \\ 0.35 \\ 0.03 \\ 0.05 \\ 0.04 \\ 0.04 \\ 0.1 \\ 0.43 \\ 0.03 \\ 1.37 \\ 0.17 \\ 0.02 \\ 0.31 \\ 1.99 \\ 0.21 \\ 0.61 \\ 0.53 \\ 0.19 \\ 0.55 \\ 0.11 \\ 0.72 \\ 0.02 \\ 0.01 \\ 0.89 \\ 0.42 \\ 0.48 \\ 0.19 \\ 0.48 \\ 0.12 \\ 0.03 \\ 0.34 \\ 0.03 \\ 0.02 \\ 1.41 \\ 0.31 \\ 0.06 \\ 0.1 \\ 0.13 \\ 0.13 \\ 6.51 \\ 0.03 \\ 2.26 \\ 0.03 \\ 0.09 \\ 0.33 \\ 0.21 \\ 0.03 \\ 0.14 \\ 0.54 \\ 0.05 \\ 0.1 \\ 0.02 \\ 0.11 \\ 0.02 \\ 0.1 \\ 0.02 \\ 0.02 \\ 0.12 \\ 1.82 \\ 0.03 \\ 0.1 \\ 0.36 \\ 0.32 \\ 0.02 \\ 0.02 \\ 0.23 \\ 0.03 \\ 0.02 \\ 0.26 \\ 0.95 \\ 0.18 \\ 0.18 \\ 0.27 \\ 0.57 \\ 0.28 \\ 0.12 \\ 0.47 \\ 0.45 \\ 0.29 \\ 0.18 \\ 0.37 \\ 0.03 \\ 0.17 \\ 0.02 \\ 0.02 \\ 0.04 \\ 0.14 \\ 0.72 \\ 0.2 \\ 0.28 \\ 0.15 \\ 0.02 \\ 0.12 \\ 0.51 \\ 0.03 \\ 1.16 \\ 0.04 \\ 2.64 \\ 0.02 \\ 2.59 \\ 1.07 \\ 0.11 \\ 0.02 \\ 0.02 \\ 0.16 \\ 1.41 \\ 0.52 \\ 0.4 \\ 0.34 \\ 0.02 \\ 0.28 \\ 0.18 \\ 0.03 \\ 0.14 \\ 5.45 \\ 1.82 \\ 0.02 \\ 0.03 \\ 0.15 \\ 0.02 \\ 0.67 \\ 0.26 \\ 0.05 \\ 0.15 \\ 0.05 \\ 0.26 \\ 0.02 \\ 0.11 \\ 0.14 \\ 0.02 \\ 0.03 \\ 0.3 \\ 0.16 \\ 0.34 \\ 2.18 \\ 0.04 \\ 0.02 \\ 0.02 \\ 0.96 \\ 0.08 \\ 0.11 \\ 0.56 \\ 0.67 \\ 0.02 \\ 0.11 \\ 0.17 \\
};   

       \addplot + [
  mark = *,
  boxplot,
  black
  ]
  table [row sep = \\, y index = 0] {
    data \\
0.02 \\ 0.03 \\ 0.03 \\ 0.04 \\ 0.03 \\ 0.04 \\ 0.04 \\ 0.06 \\ 0.03 \\ 0.1 \\ 0.14 \\ 0.03 \\ 0.13 \\ 0.12 \\ 0.01 \\ 0.3 \\ 0.07 \\ 0.05 \\ 0.02 \\ 0.18 \\ 0.03 \\ 0.12 \\ 0.03 \\ 0.04 \\ 0.04 \\ 0.02 \\ 0.03 \\ 0.21 \\ 0.18 \\ 0.11 \\ 0.03 \\ 0.02 \\ 0.03 \\ 0.15 \\ 0.32 \\ 0.01 \\ 0.05 \\ 0.03 \\ 0.01 \\ 0.02 \\ 0.03 \\ 0.05 \\ 0.22 \\ 0.01 \\ 0.07 \\ 0.06 \\ 0.03 \\ 0.02 \\ 0.89 \\ 0.08 \\ 0.02 \\ 0.02 \\ 0.03 \\ 0.11 \\ 0.03 \\ 0.13 \\ 0.02 \\ 0.03 \\ 0.06 \\ 0.12 \\ 0.05 \\ 0.35 \\ 0.75 \\ 0.04 \\ 0.04 \\ 0.32 \\ 0.05 \\ 0.02 \\ 0.85 \\ 0.08 \\ 0.08 \\ 0.13 \\ 0.15 \\ 0.02 \\ 0.05 \\ 0.14 \\ 0.11 \\ 0.11 \\ 0.04 \\ 0.25 \\ 0.17 \\ 0.53 \\ 0.56 \\ 0.03 \\ 0.03 \\ 0.02 \\ 0.5 \\ 0.03 \\ 0.1 \\ 0.12 \\ 0.2 \\ 0.03 \\ 0.14 \\ 0.14 \\ 0.03 \\ 0.12 \\ 0.28 \\ 0.05 \\ 0.07 \\ 0.15 \\ 0.02 \\ 0.06 \\ 0.14 \\ 0.05 \\ 0.04 \\ 0.01 \\ 0.02 \\ 0.03 \\ 0.06 \\ 0.14 \\ 0.02 \\ 0.03 \\ 0.07 \\ 1.89 \\ 0.4 \\ 0.05 \\ 0.28 \\ 0.02 \\ 0.21 \\ 0.53 \\ 0.36 \\ 1.7 \\ 0.56 \\ 0.01 \\ 0.04 \\ 0.16 \\ 0.03 \\ 0.1 \\ 0.12 \\ 0.04 \\ 0.37 \\ 0.77 \\ 0.03 \\ 0.08 \\ 0.15 \\ 0.07 \\ 0.04 \\ 0.02 \\ 0.16 \\ 0.06 \\ 0.03 \\ 0.85 \\ 0.05 \\ 0.43 \\ 0.03 \\ 0.05 \\ 0.06 \\ 0.13 \\ 0.02 \\ 0.05 \\ 0.03 \\ 0.02 \\ 0.13 \\ 0.02 \\ 0.02 \\ 0.38 \\ 0.02 \\ 0.04 \\ 0.18 \\ 0.33 \\ 0.01 \\ 0.27 \\ 0.15 \\ 0.03 \\ 0.29 \\ 0.19 \\ 0.28 \\ 0.02 \\ 0.31 \\ 0.07 \\ 0.03 \\ 0.02 \\ 0.03 \\ 1.68 \\ 0.03 \\ 0.03 \\ 0.14 \\ 0.14 \\ 0.06 \\ 0.19 \\ 0.06 \\ 0.32 \\ 0.01 \\ 0.04 \\ 0.03 \\ 0.03 \\ 0.05 \\ 0.03 \\ 0.1 \\ 0.04 \\ 0.02 \\ 0.22 \\ 0.03 \\ 0.17 \\ 0.03 \\ 0.04 \\ 0.09 \\ 0.02 \\ 0.04 \\ 0.03 \\ 0.15 \\ 0.01 \\ 0.03 \\ 0.06 \\ 0.07 \\ 0.04 \\ 0.03 \\ 0.13 \\ 0.04 \\ 0.02 \\ 0.23 \\ 0.05 \\ 0.17 \\ 0.21 \\ 0.14 \\ 0.03 \\ 0.03 \\ 0.04 \\ 0.02 \\ 0.15 \\ 0.03 \\ 0.23 \\ 0.22 \\ 0.03 \\ 0.02 \\ 0.14 \\ 0.06 \\ 0.03 \\ 0.01 \\ 0.07 \\ 0.02 \\ 0.86 \\ 0.29 \\ 0.35 \\ 2.31 \\ 0.09 \\ 0.03 \\ 0.06 \\ 0.03 \\ 0.06 \\ 0.03 \\ 0.03 \\ 0.08 \\ 0.03 \\ 0.04 \\ 0.03 \\ 0.29 \\ 0.1 \\ 0.02 \\ 0.08 \\ 0.09 \\ 0.03 \\ 0.07 \\ 0.54 \\ 0.04 \\ 0.24 \\ 0.33 \\ 0.03 \\ 0.08 \\ 0.14 \\ 0.15 \\ 0.03 \\ 0.01 \\ 0.1 \\ 0.04 \\ 0.06 \\ 0.32 \\ 0.12 \\ 0.76 \\ 0.02 \\ 0.37 \\ 0.25 \\ 0.02 \\ 0.12 \\ 0.04 \\ 0.02 \\ 0.04 \\ 2.29 \\ 0.03 \\ 0.05 \\ 1.81 \\ 0.03 \\ 0.02 \\ 0.02 \\ 0.1 \\ 0.04 \\ 0.03 \\ 0.05 \\ 0.05 \\ 0.1 \\ 0.05 \\ 0.03 \\ 0.03 \\ 0.03 \\ 0.18 \\ 0.27 \\ 0.13 \\ 0.03 \\ 0.12 \\ 0.22 \\ 0.06 \\ 0.04 \\ 0.05 \\ 0.01 \\ 0.03 \\
};  

       \addplot + [
  mark = *,
  boxplot,
  black
  ]
  table [row sep = \\, y index = 0] {
    data \\
0.02 \\ 0.02 \\ 0.05 \\ 0.14 \\ 0.03 \\ 0.03 \\ 0.03 \\ 0.12 \\ 0.01 \\ 0.04 \\ 0.03 \\ 0.16 \\ 0.02 \\ 0.03 \\ 0.79 \\ 1.75 \\ 0.04 \\ 0.15 \\ 0.08 \\ 0.02 \\ 0.05 \\ 1.74 \\ 0.24 \\ 0.76 \\ 0.02 \\ 0.2 \\ 5.83 \\ 0.03 \\ 0.04 \\ 0.03 \\ 0.01 \\ 0.19 \\ 0.1 \\ 0.03 \\ 0.03 \\ 0.02 \\ 0.03 \\ 0.02 \\ 0.03 \\ 0.03 \\ 0.04 \\ 0.02 \\ 0.15 \\ 0.03 \\ 0.03 \\ 0.09 \\ 0.02 \\ 0.04 \\ 0.03 \\ 0.72 \\ 0.69 \\ 0.03 \\ 0.03 \\ 0.03 \\ 0.03 \\ 0.03 \\ 0.07 \\ 0.04 \\ 0.51 \\ 0.16 \\ 0.05 \\ 0.04 \\ 0.02 \\ 0.04 \\ 0.04 \\ 0.02 \\ 0.03 \\ 0.05 \\ 0.02 \\ 0.02 \\ 0.03 \\ 1.41 \\ 0.03 \\ 0.03 \\ 0.02 \\ 0.08 \\ 0.04 \\ 0.03 \\ 1.46 \\ 0.18 \\ 0.04 \\ 0.03 \\ 0.19 \\ 0.91 \\ 0.03 \\ 0.04 \\ 0.02 \\ 0.14 \\ 0.03 \\ 0.03 \\ 26.04 \\ 0.01 \\ 0.03 \\ 0.04 \\ 0.02 \\ 0.01 \\ 0.03 \\ 3.31 \\ 0.03 \\ 3.9 \\ 0.08 \\ 0.03 \\ 0.12 \\ 0.03 \\ 0.04 \\ 0.03 \\ 0.03 \\ 0.94 \\ 0.04 \\ 0.02 \\ 0.04 \\ 0.02 \\ 0.04 \\ 2.28 \\ 0.03 \\ 0.03 \\ 0.03 \\ 0.85 \\ 0.22 \\ 0.56 \\ 0.03 \\ 0.03 \\ 0.03 \\ 0.03 \\ 0.09 \\ 5.17 \\ 0.08 \\ 0.05 \\ 0.03 \\ 0.02 \\ 5.6 \\ 0.04 \\ 0.03 \\ 0.03 \\ 0.03 \\ 0.03 \\ 0.17 \\ 1.08 \\ 0.04 \\ 0.14 \\ 0.16 \\ 0.02 \\ 0.01 \\ 0.02 \\ 0.81 \\ 0.03 \\ 3.62 \\ 9.32 \\ 0.02 \\ 0.02 \\ 0.02 \\ 0.03 \\ 0.92 \\ 0.03 \\ 0.02 \\ 0.02 \\ 0.04 \\ 0.04 \\ 0.06 \\ 0.04 \\ 0.09 \\ 0.14 \\ 0.03 \\ 0.03 \\ 0.47 \\ 0.02 \\ 0.11 \\ 0.01 \\ 0.03 \\ 0.03 \\ 0.06 \\ 0.48 \\ 0.66 \\ 0.02 \\ 0.02 \\ 0.02 \\ 0.16 \\ 0.28 \\ 1.7 \\ 0.04 \\ 0.35 \\ 0.5 \\ 0.03 \\ 0.16 \\ 0.03 \\ 0.04 \\ 0.05 \\ 0.03 \\ 0.02 \\ 3.92 \\ 1.77 \\ 0.02 \\ 0.02 \\ 0.69 \\ 0.55 \\ 0.03 \\ 0.03 \\ 25.42 \\
};  

  \end{axis}
\end{tikzpicture}
\caption{Commit processing time distribution in four projects.}
\label{fig:executionTime}
\end{figure}

A box plot of distribution of execution time per project is presented in~\Cref{fig:executionTime}.
The maximum median time of \toolName on Java projects is equal to 0.05 seconds, and on Kotlin projects, 0.16 seconds, which, we believe, is acceptable in our task.
The results of measurements are available online: \url{https://zenodo.org/record/4905639}.

\section{Conclusions and future work}

In this paper, we presented \toolName, a plugin for IntelliJ IDEA that integrates information about refactorings in diffs in the IDE, auto folds them, and shows hints with short descriptions of the refactorings.
To detect refactorings in code changes in Java and Kotlin, \toolName relies on RefactoringMiner and kotlinRMiner respectively.
\toolName allows users to review refactoring changes separately from other changes in pull requests and in specific commits as well as to explore the refactoring history of objects.
The evaluation shows that the median execution time of \toolName is less than 0.2 seconds, which proves that extracting and visualizing information about refactorings in an IDE does not interrupt developers' workflow.

In our ongoing work on improving \toolName, we plan to conduct a case study involving software developers to investigate the plugin's usefulness.
Also, we plan to continue working on kotlinRMiner: support other refactoring types and collect representative dataset containing refactorings in Kotlin for its evaluation.

\balance

\bibliographystyle{IEEEtran}
\bibliography{refactorinsight}

\end{document}